\documentclass[aps,prl,reprint,superscriptaddress,amsmath,amssymb,longbibliography,nobibnotes]{revtex4-2}
\usepackage[T1]{fontenc}
\usepackage{lmodern}
\usepackage{graphicx}
\usepackage{bm}
\usepackage{braket}
\usepackage{booktabs}
\usepackage{tikz}
\usetikzlibrary{arrows.meta,calc}
\usepackage{pgfplots}
\pgfplotsset{compat=1.18}
\usepackage[colorlinks=true,linkcolor=blue,citecolor=blue,urlcolor=blue]{hyperref}
\usepackage{float}

\definecolor{paBlue}{RGB}{24,81,132}
\definecolor{paOrange}{RGB}{193,91,31}
\definecolor{paGreen}{RGB}{30,118,108}
\pgfplotsset{paaxis/.style={width=\columnwidth,height=4.0cm,
  tick label style={font=\footnotesize},label style={font=\footnotesize},
  legend style={font=\scriptsize,draw=none,fill=white},
  axis line style={black!70},tick style={black!70},
  grid=major,grid style={black!10},line width=0.55pt}}

\begin{document}

\title{Search for the $^{229}$Pa Parity Doublet via Internal-Conversion Spectroscopy at FRIB}
\author{Silviu-Marian Udrescu}
\email{sudrescu@princeton.edu}
\affiliation{\mbox{Department of Physics, Princeton University, Princeton, NJ 08544, USA}}
\affiliation{\mbox{Department of Physics and Astronomy, Johns Hopkins University, Baltimore, MD 21218, USA}}

\author{Aiden Boyer}
\affiliation{\mbox{Department of Physics and Astronomy, Michigan State University, East Lansing, MI 48824, USA}}
\affiliation{\mbox{Facility for Rare Isotope Beams, East Lansing, MI 48824, USA}}

\author{Lan Cheng}
\affiliation{\mbox{Department of Chemistry, Johns Hopkins University, Baltimore, MD 21218, USA}}

\author{Ronald Fernando Garcia Ruiz}
\affiliation{\mbox{Massachusetts Institute of Technology, Cambridge, MA 02139, USA}}

\author{Phay J. Ho}
\affiliation{\mbox{Chemical Sciences and Engineering Division, Argonne National Laboratory, Lemont, IL 60439, USA}}

\author{Ryan Ringle}
\affiliation{\mbox{Department of Physics and Astronomy, Michigan State University, East Lansing, MI 48824, USA}}
\affiliation{\mbox{Facility for Rare Isotope Beams, East Lansing, MI 48824, USA}}

\author{Jaideep T. Singh}
\affiliation{\mbox{Department of Physics and Astronomy, Michigan State University, East Lansing, MI 48824, USA}}
\affiliation{\mbox{Facility for Rare Isotope Beams, East Lansing, MI 48824, USA}}

\author{Patrick R. Stollenwerk}
\affiliation{\mbox{Department of Physics and Astronomy, Michigan State University, East Lansing, MI 48824, USA}}
\affiliation{\mbox{Facility for Rare Isotope Beams, East Lansing, MI 48824, USA}}

\author{Georg Tiebel}
\affiliation{\mbox{Department of Physics and Astronomy, Michigan State University, East Lansing, MI 48824, USA}}
\affiliation{\mbox{Facility for Rare Isotope Beams, East Lansing, MI 48824, USA}}

\author{Margaret J. Tseng}
\affiliation{\mbox{Department of Physics and Astronomy, Michigan State University, East Lansing, MI 48824, USA}}
\affiliation{\mbox{Facility for Rare Isotope Beams, East Lansing, MI 48824, USA}}

\begin{abstract}
The radioisotope $^{229}$Pa is predicted to possess a nearly degenerate ground-state parity doublet. Previous $\gamma$-ray and conversion-electron spectroscopy suggests a splitting of $\Delta E = (60\pm 50)~\mathrm{eV}$, although the doublet's existence remains unconfirmed. Such a small splitting would give $^{229}$Pa an exceptionally large enhancement of symmetry-violating nuclear properties. However, the splitting, and with it the sensitivity of $^{229}$Pa to CP-violating physics, remains uncertain by orders of magnitude. We propose an experiment to measure this splitting with sub-eV precision within a few days of measurement time by using delayed-coincidence internal-conversion spectroscopy of $^{229}$Pa ions produced by the electron-capture decay of $^{229}$U$^{+}$ ions stored in a Penning trap at the Facility for Rare Isotope Beams (FRIB). This measurement would provide the key nuclear-structure input needed to establish $^{229}$Pa as a leading candidate for next-generation atomic and molecular electric-dipole-moment (EDM) experiments.

\end{abstract}
\maketitle

\section{Introduction}

The observed dominance of matter over antimatter in the Universe~\cite{Planck2020} remains one of the central open problems connecting particle physics and cosmology. Generating such an asymmetry dynamically requires, among Sakharov's conditions, the violation of charge conjugation and parity (CP) symmetries~\cite{Sakharov1967}. The CP violation contained in the Standard Model (SM), namely the complex phase of the Cabibbo--Kobayashi--Maskawa matrix, falls short of what is required by many orders of magnitude~\cite{Gavela1994,Morrissey2012}. This strongly suggests additional sources of CP violation beyond the SM. Permanent electric dipole moments (EDMs) of nucleons, atoms, and molecules are among the most sensitive probes of such physics~\cite{Safronova2018,Chupp2019,udrescu2026probing}. A nonzero EDM violates parity (P) and time-reversal (T) symmetry, and therefore CP via the CPT theorem. Because SM contributions lie far below present experimental reach, EDM searches are effectively background-free probes of new physics at mass scales that can exceed those accessible directly at colliders~\cite{Pospelov2005,Engel2013,Safronova2018,Chupp2019,udrescu2026probing}.
 
In diamagnetic atoms and molecules, CP-violating phenomena appear primarily through the
nuclear Schiff moment $S$, the residual P,T-odd moment that survives electronic
screening of the nuclear EDM~\cite{Schiff1963,Sushkov1984}, given in the lab frame, to first order in perturbation theory, by \cite{engel2025nuclear}: 
\begin{equation}
  S = -\sum_{k}
  \frac{\langle 0|\hat{S}_z|k\rangle\langle k|\hat{V}_{PT}|0\rangle}{\Delta E_k}
  + \mathrm{c.c.},
  \label{eq:schiff}
\end{equation}
where $\hat{S}_z$ is the intrinsic body-frame Schiff moment, $\ket{0}$ and $\ket{k}$ are the ground-state nuclear wavefunction and the wavefunctions of excited states with opposite parity and the same total angular momentum as the ground state, respectively, while $\Delta E_k$ is their energy difference. Here, $\hat{V}_{PT}$ is the P,T-odd nucleon--nucleon
interaction containing the possible beyond-SM physics of interest. In near-spherical nuclei such as $^{199}$Hg, which currently sets the best limits on nuclear CP-violating phenomena~\cite{Graner2016}, such opposite-parity states lie at MeV energies, which suppresses $S$ and the induced atomic and molecular EDMs~\cite{engel2025nuclear}.
 
The situation changes significantly in nuclei with octupole (pear-shaped) deformation \cite{Butler1996,Gaffney2013,ArrowsmithKron2024}. Intrinsic reflection asymmetry gives rise to nearly degenerate parity doublets, whose splitting, $\Delta E$, can be significantly smaller than typical nuclear energies, on the order of tens of keV~\cite{ArrowsmithKron2024,Haxton1983,Auerbach1996,Spevak1997,Engel2000}. In this case, the laboratory-frame Schiff moment is dominated by a single term in Eq.~\eqref{eq:schiff}, corresponding to the opposite-parity partner of the ground state. Schiff moments of octupole-deformed nuclei are consequently predicted to exceed those of near-spherical nuclei by two to three orders of magnitude~\cite{Auerbach1996,Spevak1997,Engel2000,Dobaczewski2018,engel2025nuclear}. As a result of this, significant efforts have emerged towards using such octupole-deformed isotopes in searches for fundamental symmetry violations~\cite{Parker2015,ArrowsmithKron2024,udrescu2026probing}. 

Very recently, molecules containing such nuclei have also become a realistic platform in searches for new CP-violating physics~\cite{ArrowsmithKron2024,jadbabaie2026radioactive,udrescu2026probing}. Closely spaced opposite-parity molecular levels can allow the molecule to be fully polarized in modest laboratory electric fields, which further enhances P,T-odd energy shifts by two to three orders of magnitude relative to atoms~\cite{Safronova2018}. This principle underlies the leading electron-EDM limits from ThO~\cite{ACME2018} and HfF$^{+}$~\cite{Roussy2023} and ongoing Schiff-moment searches with TlF~\cite{Grasdijk2021}. The prospect of combining nuclear and molecular enhancements has driven a rapidly growing program on radioactive molecules~\cite{Flambaum2020,Isaev2010,ArrowsmithKron2024,jadbabaie2026radioactive,udrescu2026probing}, highlighted by the first laser-spectroscopic study of a short-lived, radioactive molecule, RaF~\cite{GarciaRuiz2020,udrescu2021isotope,udrescu2024precision,wilkins2025observation,wilkins2026ionization,Conn2026ColdRadioactiveMolecules}. However, the projected reach of any such experiment depends on the Schiff moment, $S$, of the heavy nucleus in the molecule, and therefore on $\Delta E$. Consequently, measuring this nuclear-structure property with sufficient precision is a prerequisite for selecting among candidate species~\cite{ArrowsmithKron2024}.
 
$^{229}$Pa is of particular interest among such candidates. Chasman predicted that
$^{229}$Pa, at the edge of the octupole region, would host a nearly degenerate
ground-state parity doublet~\cite{Chasman1980}. In 1982, Ahmad~\textit{et al.} reported a $5/2^{+}$--$5/2^{-}$ ground-state doublet split by only
$(220\pm50)$~eV, based on the electron-capture (EC) decay of
$^{229}$U~\cite{Ahmad1982}. Haxton and Henley immediately recognized the potential of this exceptionally small energy splitting to enhance T-violating nuclear moments by orders of magnitude~\cite{Haxton1983}. Later model calculations supported the doublet interpretation~\cite{Leander1988}, but subsequent experimental studies questioned the original level assignments~\cite{grafen1991does}. A more recent re-analysis of high-resolution $\gamma$-ray and conversion-electron data indicated an energy splitting of $\Delta E = (60\pm50)$~eV, too uncertain to confirm the existence of the doublet \cite{Ahmad2015}. Theoretical inputs vary widely: Schiff-moment calculations have assumed energy splittings of $\sim$5~keV~\cite{Spevak1997}, whereas recent beyond-mean-field calculations place the $5/2^{-}$ bandhead near 16~keV~\cite{Zhou2025}. Given that
$S \propto 1/\Delta E$, the resulting sensitivities span several orders of magnitude. 

Here, we propose a direct search for this parity doublet in $^{229}$Pa with a target uncertainty below $1~\mathrm{eV}$, by leveraging the unique opportunities offered by the Facility for Rare Isotope Beams (FRIB)~\cite{Glasmacher2024}. If confirmed, an energy splitting on the order of tens of eV could result in a Schiff moment in $^{229}$Pa approximately $10^{3}$ times larger than that of $^{225}$Ra and up to $10^{6}$ times larger than that of $^{199}$Hg~\cite{ArrowsmithKron2024,singh2019new}. This would allow future EDM experiments to explore a wide, unexplored range in parameter space for new CP-violating physics, at energy scales of hundreds and even thousands of TeV~\cite{ArrowsmithKron2024,jadbabaie2026radioactive,udrescu2026probing}.

\section{Measurement Strategy}

If the parity doublet in $^{229}$Pa lies within the expected energy range of tens to hundreds of eV~\cite{Ahmad2015}, and the excited state has spin--parity $5/2^{-}$, it would decay to the $5/2^{+}$ ground state via an $E1$ transition, with internal conversion (IC) expected to dominate whenever energetically allowed~\cite{Dragoun1993,Grechukhin2000}. During IC, the nuclear excitation energy is transferred to a bound electron, resulting in its ejection from the atom or ion. Consequently, $\Delta E$ corresponds to the sum of the kinetic energy of the emitted electron and the energy required to access the corresponding ionization channel, including any excitation of the residual ion~\cite{vonderWense2016,Tiedau2024,Seiferle2019}. Thus, electron spectroscopy, combined with knowledge of the initial and final electronic states, enables a direct determination of the parity-doublet energy splitting, $\Delta E$.

It is worth mentioning that this approach has a well-established precedent: the detection of IC electrons enabled both the first direct observation of the $^{229\mathrm{m}}$Th isomer~\cite{vonderWense2016} and the first direct determination of its excitation energy, $8.28(17)$~eV~\cite{Seiferle2019}. These landmark measurements paved the way for the subsequent precision measurements of this nuclear transition ~\cite{Tiedau2024}. However, applying this approach to $^{229}$Pa requires a different experimental strategy, as outlined below.

\begin{figure*}[t]
    \centering
    \includegraphics[width=\textwidth]{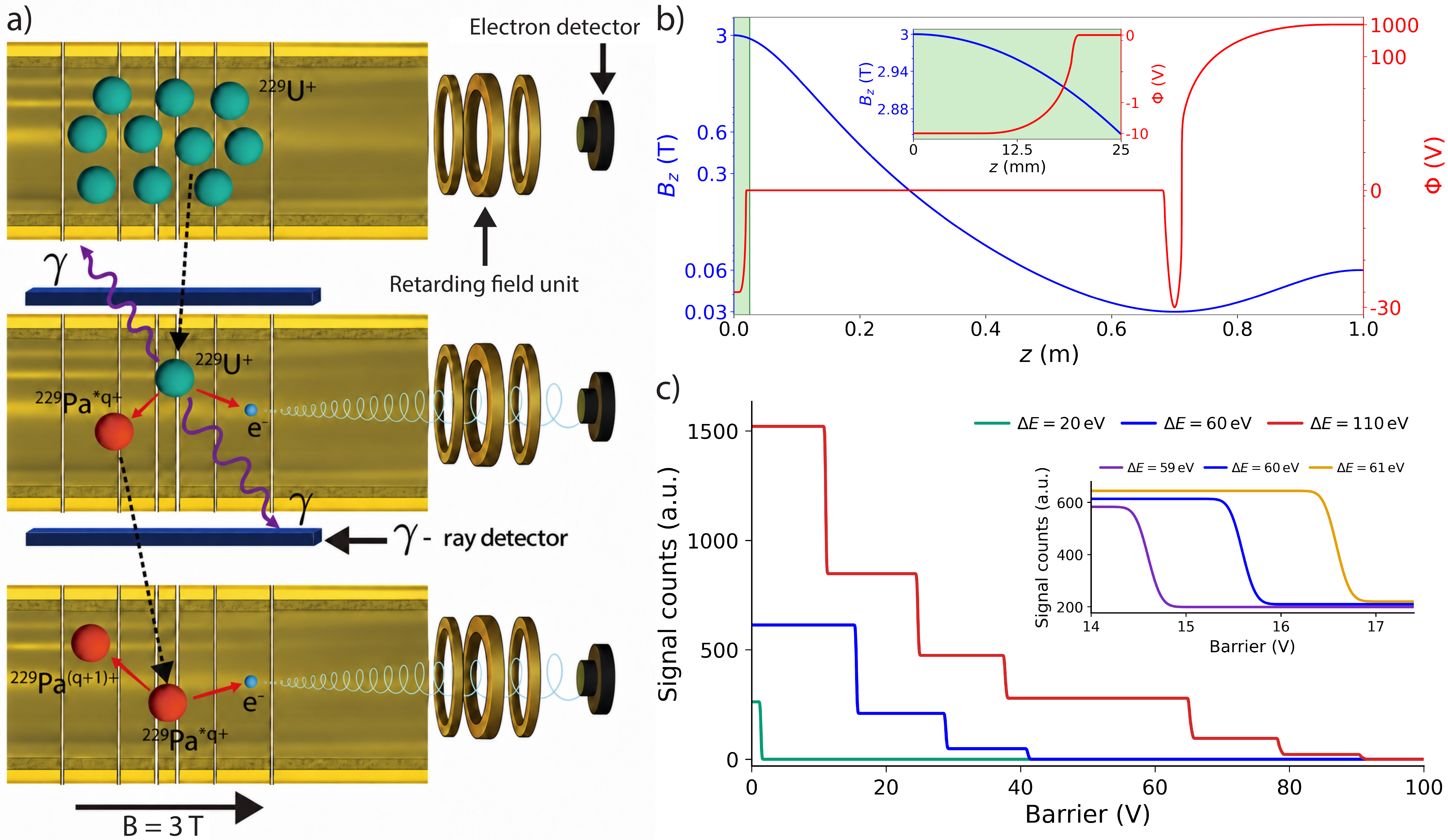}
    \caption{\textbf{Proposed experimental setup and expected signal.}  \textbf{(a)}~Measurement principle (electrodes shown in yellow, not to scale).  \textit{Top:} a cloud of $\sim 10^{4}$ $^{229}$U$^{+}$ ions (green) is stored and  cooled in a cylindrical Penning trap in a magnetic field $B = 3~\mathrm{T}$, with a retarding-field unit and a microchannel-plate (MCP) electron detector placed on the magnet axis.  \textit{Middle:} $^{229}$U$^{+}$ decays by electron capture ($T_{1/2} = 57.8~\mathrm{min}$,  80\% branch) to $^{229}$Pa (orange). The nuclear de-excitation and the resulting atomic relaxation cascade emit X-rays, $\gamma$~rays, and electrons within  $\lesssim 1~\mathrm{ns}$, leaving a Pa$^{q+}$ ion; detection of these prompt particles in the surrounding photon detectors and/or the MCP serves as the trigger. In $\sim 10\%$ of the decay events, the upper member of the candidate parity doublet is populated (if it exists). \textit{Bottom:} the excited $^{229}$Pa$^{q+}$ ion decays to the nuclear ground state by IC, emitting an electron and leaving  Pa$^{(q+1)+}$. The IC electron is guided along the magnetic field through the retarding unit to the MCP; the distribution of delays relative to the trigger yields the lifetime of the excited state, and the retarding voltage at which it is suppressed provides its kinetic energy $E_{\mathrm{kin}}$, from which  $\Delta E = E_{\mathrm{kin}} + I_q$. A second MCP (not shown), without a retarding field unit, will be positioned at the other end of the trap. \textbf{(b)}~On-axis magnetic field $B_z$ (blue, left axis) and electrostatic potential $\Phi$ (red, right axis, nonlinear scale) as a function of the distance $z$ from the trap center; the shaded area marks the trap region, shown enlarged in the inset. The field decreases from $3~\mathrm{T}$ at the trap to  $\sim 0.03~\mathrm{T}$ at the analyzing plane ($z \approx 0.7~\mathrm{m}$), where the  retarding potential is applied (here $20~\mathrm{V}$ relative to the emission potential), before the electrons are post-accelerated to $1~\mathrm{kV}$ onto the MCP at $z \approx 1~\mathrm{m}$ ($B \approx 0.06~\mathrm{T}$). \textbf{(c)}~Simulated integral spectra of detected IC electrons as a function of the barrier voltage for $\Delta E = 20~\mathrm{eV}$ (green), $60~\mathrm{eV}$ (blue), and $110~\mathrm{eV}$ (red). Each descending step corresponds to IC in a given Pa charge state $q$, at $E_{\mathrm{kin}} = \Delta E - I_q$, with ionization energies $I_q$ from Ref.~\cite{NISTASD} and step heights set by the calculated charge-state distribution following the EC event from $^{229}$U. The curves are scaled individually for display; the absolute rates increase by about an order of magnitude for every $\sim 50~\mathrm{eV}$ increase in  $\Delta E$ (see main text). \textit{Inset:} Pa$^{4+}$ step for $\Delta E = 59,~60,~\mathrm{and}~61~\mathrm{eV}$, illustrating the  sensitivity to $1~\mathrm{eV}$ changes in $\Delta E$. The curves are shifted vertically for clarity.}
    \label{figure_setup}
\end{figure*}

The proposed experiment will begin with trapped $^{229}$U$^+$ ions inside a Penning trap (Fig. \ref{figure_setup}a). With a half-life of $57.8~\mathrm{min}$, $^{229}$U decays through EC to $^{229}$Pa with an approximately $80\%$ branching ratio~\cite{Ahmad2015}. The candidate excited member of the parity doublet may be populated directly or through cascades from higher nuclear levels, with an estimated probability of $\sim 10\%$ per EC decay~\cite{Ahmad2015}. During the EC process, an inner-shell electron vacancy is typically formed in the Pa daughter, resulting in the emission of X-rays and Auger electrons~\cite{Bambynek1972}. Shake-off and internal conversion of higher-energy nuclear transitions can eject additional electrons. These electrons and X-rays can be used as a trigger for our experiment. Furthermore, these processes generate Pa ions with a final charge $q$ that varies from event to event, with a broad distribution that peaks near $q = 11$ (see Supplemental Material).

Because $\Delta E$ may be comparable to the binding energies of the outer electrons in the daughter ions, the nuclear decay is sensitive to the electronic structure of Pa~\cite{Grechukhin2000}. Consequently, the IC spectrum may contain multiple lines associated with different charge states and electronic ionization channels. For ions in their electronic ground state, direct IC becomes energetically forbidden when the ionization energy exceeds $\Delta E$~\cite{Seiferle2019}. The charge-dependent line positions and conversion thresholds would provide a characteristic signature of the nuclear transition. A consistent value of $\Delta E$ inferred from several electronic channels, combined with coincidences associated with the prompt photons and electrons emitted after the EC event, would provide a clear signature while strongly suppressing backgrounds. 

\section{Experimental Details}

Below, we describe in detail the proposed experiment, which is designed to search near the reported parity-doublet splitting of $(60\pm50)~\mathrm{eV}$~\cite{Ahmad2015}, with a target uncertainty below $1~\mathrm{eV}$ in a few days of measurement time. The experiment starts with trapping $\sim10^4$ singly charged $^{229}$U$^+$ ions inside a Penning trap (Fig. \ref{figure_setup}a, top). At FRIB, a rate of $>5 \times 10^5$ $^{229}$U$^+$ ions per second, at an energy of tens of keV, is expected to be delivered to our experimental setup~\cite{FRIBRates}. The trap will be floated at a voltage chosen to facilitate the capture of the incoming ions with high efficiency. To provide the axial access required for electron detection, as described below, a cylindrical Penning trap with open end caps and an inner diameter of $2~\mathrm{cm}$ will be utilized~\cite{Gabrielse1989}. The trap will consist of seven independently biased electrodes (Fig. \ref{figure_setup}a), which allow the axial potential to be adjusted from approximately harmonic during capture and cooling to nearly flat (box-like) during the actual measurement, while retaining confining barriers at the ends. This latter configuration minimizes variations in the electron emission potential across the ion cloud, reducing broadening of the measured energy spectrum. 

The ions are cooled to $\sim 4~\mathrm{K}$ with pulsed cryogenic He buffer gas, while a rotating-wall field applied to azimuthal segments of the central electrode compresses the cloud to a diameter of $\sim 0.5~\mathrm{mm}$ and a length of $\sim 1~\mathrm{cm}$~\cite{Huang1997}. The buffer gas is then removed to limit electron scattering and charge exchange of the daughter ions. Capture, cooling, and cloud preparation are expected to take $\lesssim 1~\mathrm{s}$, which is small when compared to the trap cycle of $1~\mathrm{min}$ described below.

Each trap cycle will be 1 min long. During this period, approximately 95 EC decays (${\sim}5.7\times10^3~\mathrm{h^{-1}}$) will produce $^{229}$Pa (Fig.~\ref{figure_setup}a, middle). At the end of each trap cycle, the remaining $\sim10^4$ $^{229}$U$^+$ ions will be extracted and counted, with parent and daughter contributions distinguished in the ion readout. Together with the known parent half-life and EC branching fraction, this count will normalize the initial parent population. The measurement trap will then be ready for the next batch. The retarding barrier is kept fixed for 20 consecutive trap cycles, which we refer to below as a $20~\mathrm{min}$ measurement cycle. Each measurement cycle thus samples ${\sim}1.9\times10^3$ EC decays. 

During each $1~\mathrm{min}$ trap cycle, newly delivered $^{229}$U ions will be accumulated in a separate storage ion trap in parallel with electron acquisition. Even with a loading rate as low as 200 ions per second, corresponding to $<0.1\%$ of the assumed delivered rate, more than $10^4$ $^{229}$U ions would be accumulated after $1~\mathrm{min}$, accounting for radioactive decay during accumulation. The accumulated sample will be purified before injection to suppress decay daughters produced during storage. Parallel accumulation allows repeated measurements with a high duty factor, with dead time determined primarily by ion transfer, buffer-gas cooling, and removal of the buffer gas before acquisition.

During each EC decay of $^{229}$U, the atomic relaxation cascade produces X rays and Auger electrons on short time scales ($<1~\mathrm{ns}$), leaving the Pa daughter in a distribution of charge states. Detected prompt events provide the trigger and timing reference for the measurement. We estimate that the candidate excited parity-doublet state is populated in $\sim10\%$ of the EC decays to $^{229}$Pa~\cite{Ahmad2015}, corresponding to approximately $190$ candidate-state populations per measurement cycle. If energetically allowed, this state can subsequently decay to the ground state by emitting an IC electron (Fig. \ref{figure_setup}a, bottom). Its lifetime will depend on the charge state of the Pa daughter and on the overlap of its electronic wavefunction with the Pa nucleus, and it is expected to be in the range of a few to a few hundred microseconds~\cite{Dragoun1993,grechukhin2000studying}. However, the timing analysis will allow us to cover delays relative to the trigger from hundreds of nanoseconds into the millisecond range, with the search window adjusted to the observed signal and instrumental response. A time distribution analysis will allow us to experimentally extract the lifetime of the excited parity-doublet state for the different Pa charge states present. For the initial population of $10^4$ ions, the mean interval between $^{229}$U decays is approximately 0.5~s throughout each trap cycle, providing substantial separation between parent decay events for submillisecond coincidence windows. The residual pile-up of decays within a window is discussed below.

The photons produced following the EC event include Pa K X rays at
92--112~keV and $\gamma$ rays from the nuclear de-excitation of $^{229}$Pa~\cite{Ahmad2015}. These can be detected by placing detectors around the trap in a barrel-like configuration~\cite{cooper2012gamma}. Given the ion cloud size, an array subtending $\pi$ sr relative to the trap center is feasible~\cite{cooper2012gamma}, providing $\sim 25\%$ geometrical coverage for isotropic photon emission. Since several photons are emitted per decay, the probability of detecting at least one is considerably higher.

The prompt electron spectrum comprises Auger and shake-off electrons, together with conversion electrons from nuclear de-excitation cascades. These electrons can be guided along the magnetic field to two position-sensitive MCP detectors placed axially at opposite ends of the trap, where the field strength decreases away from the magnet center~\cite{Seiferle2019}. We assume $\sim 50\%$ intrinsic electron-detection efficiency for each MCP and $50\%$ directional acceptance for each arm, giving a single-electron detection probability of $25\%$ per arm. It is worth mentioning that combined photon and electron detection would improve the probability of obtaining a prompt EC trigger, but either channel can also be used independently, given that multiple prompt particles will be emitted in each relaxation cascade.

The IC electrons can be detected using the same setup as the prompt electrons, but their kinetic energies are measured with a retarding-field analyser similar to that used in $^{229\text{m}}$Th IC spectroscopy~\cite{Seiferle2019}. A tunable electrostatic barrier between the trap center and one of the MCPs (see Fig. \ref{figure_setup}a), referenced to the floating trap potential and followed by post-acceleration, transmits only electrons with sufficient longitudinal energy (Fig. \ref{figure_setup}b). Placing the barrier in a region of reduced magnetic field collimates the electrons adiabatically, so that the threshold approaches the total kinetic energy, with a residual angular broadening set by the field ratio and included in the calibrated transmission function (see Supplemental Material for details). Although the second MCP is not used for energy measurement, it significantly increases the trigger efficiency and helps reduce the background.

Scanning the barrier yields an integral spectrum, normalized to the number of EC decays sampled at each setting. Resolved conversion-electron lines will produce downward steps as the barrier height increases, with the step positions extracted by fitting the calibrated spectrometer transmission function (Fig. \ref{figure_setup}c). For an identified conversion channel, $E_{\mathrm{kin}}=\Delta E-I_q$, where $I_q$ is the ionization energy of the Pa daughter. Agreement between the values of $\Delta E$ inferred from different channels would provide a consistency check on the signal. The absolute energy scale and transmission response will be calibrated using electrons of known energy as described below. 

\subsection{Background and systematics} 

We will next consider various sources of background that can mimic our signal of interest. The 20\% $\alpha$-decay branch of $^{229}$U produces 6.3-MeV $\alpha$ particles and 110-keV $^{225}$Th recoils~\cite{Ahmad2015}, which leave the cloud volume. They will be mostly implanted in the trap electrodes, while for an  MCP located $1$ m away from the trap center, any effect due to direct hits will be negligible. Prompt atomic electrons accompanying the $\alpha$ decay may also be guided to the MCP. However, the $^{225}$Th leaves the cloud within a few tens of nanoseconds and therefore does not remain as a delayed electron source. Daughter activity from this decay, or from contaminants injected with the beam, implanted on surrounding surfaces, can produce uncorrelated electrons. Their rate and spatial distribution will be independently measured with the trap empty. Moreover, since electrons follow magnetic flux surfaces under adiabatic transport, those emitted outside the flux tube of the cloud land at larger radii on the MCP and can be removed by a position cut on the detector or a physical axial aperture.

MCP dark counts could also represent a source of accidental coincidences. We assume an intrinsic dark-count density of $0.01$ s$^{-1}$ cm$^{-2}$ and a coincidence window of $100~\mu$s. By placing the MCP in a region with a magnetic field of $\sim 0.06$ T (see Fig. \ref{figure_setup}b), a 2 mm source diameter maps to an approximately 14-mm diameter image. Thus, the expected dark-count contribution to the background is $\sim 0.26$ counts/day. We choose a $2$ mm initial diameter in order to account for recoils of the $^{229}$Pa daughter out of the initial $0.5$ mm diameter $^{229}$U cloud.

Pile-up of other decays is a further source of accidental coincidences. At an average rate of $\sim 2$ decays per second, a second decay occurs within the $100$ $\mu$s window with probability $\sim 2\times10^{-4}$. It is vetoed if its prompt cluster produces a detected K X-ray or two or more electron hits across the two MCPs, but a single detected electron without a photon mimics an IC electron.  For the calculated Pa charge distribution, this happens in $\approx 1.4\%$ of EC decays, giving a background of $0.49$ accidental counts/day.  This rate, and its prompt-electron multiplicity, will also be determined experimentally at each retarding setting from pre-trigger windows or suitably time-shifted trigger coincidences, using the same event selection and the appropriate exposure normalization.

Delayed counts also arise from the atomic cascade itself. Auger-cascade calculations for the analogous EC decay of $^{125}$I give $\sim 5$ electrons per decay with energies below $100$ eV~\cite{Pomplun2000}, overlapping the IC-electron energy range. Although these electrons are emitted promptly, those that clear the retarding barrier with little excess energy cross the analyser slowly and can arrive microseconds after the trigger. The number of such electrons arriving later than a delay $t$ falls as $t^{-2}$. Opening the coincidence window at $t_0 \approx 1~\mu$s gives an estimated $0.49$ late-prompt counts/day (see Supplemental Material). 

An important contribution to the energy uncertainty on $\Delta E$ is the calibration of the retarding barrier relative to the electron-emission potential. This can be calibrated in situ with electrons of known energy, either released from trapped electrons by controlled axial excitation or produced by photodetachment of negative ions such as Th$^{-}$. Using Th$^-$ could offer further advantages because its similar mass to $^{229}$U and $^{229}$Pa would permit tests of heavy-ion transport and cooling. In either case, the calibration will characterize the absolute energy scale and spectrometer transmission over the relevant electron-energy range, including space charge effects and angular broadening. The goal is to achieve a calibration uncertainty $<0.1$ eV. To account for temporal drifts, such calibration measurements can be interleaved with actual $^{229}$Pa measurements. 

Finally, the measured energy depends on the potential at the emission point. This has two main contributions: the applied trap potential, which can vary as a function of position, and the potential created by the other ions inside the trap. The first contribution can be reduced by adjusting the trap potential to a nearly flat central region with confining end barriers during the measurement. SIMION simulations give variations of the applied potential below $50$ meV across the cloud~\cite{Dahl2000}. For a uniform-density cloud~\cite{DubinONeil1999}, the space charge adds a spread of $\approx 2$~mV across the cloud. Overall, these contributions, together with uncertainties in the electronic removal energies, will enter the uncertainty budget for the targeted sub-eV precision in $\Delta E$.

\subsection{Signal estimate} 

The signal depends on the fraction $f_\text{IC}$ of daughters for which IC is energetically allowed. For example, for $\Delta E \approx 60$~eV, only Pa$^{q+}$ ions with $q \le 4$ can convert from their electronic ground states~\cite{NISTASD}. These fractions have been calculated in this work (see Supplemental Material). Approximately $2.7\%$ of the ions end with $q \leq 4$, so we take $f_{IC} \approx 0.027$ at $\Delta E = 60$~eV. The fraction $f_\text{IC}$ rises steeply for larger splittings, from $\approx 0.003$ at $20$ eV, where only Pa$^{+}$ and Pa$^{2+}$ convert, to $\approx 0.13$ at $110$ eV, where Pa$^{5+}$ to Pa$^{7+}$ also convert.

We trigger on either a detected Pa K X-ray or one or more prompt electrons hitting either MCP within the region onto which the ion cloud is imaged. A daughter of charge $q$ has emitted $q-1$ electrons, so IC-open daughters at $\Delta E \approx 60$~eV are accompanied by at most three prompt electrons, typically together with a K X-ray. With an intrinsic electron-detection efficiency of $50\%$ for each MCP and $50\%$ directional acceptance per arm, the probability of detecting a prompt electron in either MCP is $0.50$. Delayed IC electrons are counted only in the retarded arm, with detection probability of $0.25$. We adopt a K-X-ray detection probability of $0.15$ (assuming an intrinsic detector efficiency of $\sim 90\%$ at 100~keV and $\sim 65\%$ transmission through the trap electrodes). The prompt-trigger probability is therefore $\eta_q=1-0.85(0.50)^{q-1}$ for $q\geq2$, while the $q=1$ channel is taken to be untagged. The $\sim 190$ candidate-state populations per measurement cycle thus yield approximately $5.9$, $75$, and $407$ counts/day at $\Delta E=20$~eV, $60$~eV, and $110$~eV. Detecting Pa L X rays could further raise the trigger efficiency.

This signal should be compared with the backgrounds discussed above. For the two-MCP configuration, after spatial and timing selection, MCP dark counts contribute approximately $0.26$ counts/day, while correlated late-prompt electrons and residual pile-up contribute approximately $0.49$ counts/day each. The total expected background is therefore approximately $1.24$ counts/day, compared with approximately $75$ signal counts/day at $\Delta E=60$~eV. 

The measurement will proceed adaptively. First, we measure the delayed-electron yield at zero retarding barrier, which transmits all energetically allowed IC channels. Observation of an excess identified as IC decay of the candidate opposite-parity state would already establish the existence of the parity doublet.

After each $20$-min measurement cycle, the delayed counts from all cycles so far, sorted by the number of detected prompt electrons, are fitted with the expected spectrum: steps at $eU_q=\Delta E-I_q$ (Fig.~\ref{figure_setup}c) with heights and multiplicities set by the calculated charge-state distribution and the detection probabilities, plus background. For our simulations, the signal and background rates are free parameters, constrained by log-normal priors centered on the estimates above, with widths of a factor of 3 and 30\%, respectively. As a Pa$^{q+}$ daughter has emitted $q-1$ electrons, the multiplicity helps identify the charge state behind each step, and further suppresses the background, which is triggered mostly by high-$q$ decays. The barrier for each new cycle is chosen to maximize the expected information gain on $\Delta E$. The measurement stops when, in two consecutive cycles, the posterior holds $\ge99\%$ probability within $\pm3$~eV of its maximum, so that alternative charge-state assignments carry  $\le 1\%$ posterior probability, and its $68\%$ credible interval has a half-width $\le 1$ eV. Based on numerical simulations, the predicted average measurement times are 117.4~h, 31.4~h, and 8.7~h for $\Delta E=20$, 60, and 110~eV (see Supplemental Material). Larger samples would shorten the measurement further: with $10^5$ ions per trap cycle, which would still use less than $1\%$ of the assumed $^{229}$U$^+$ delivery rate at FRIB, the same simulations give average measurement times of 18.5~h, 4.0~h, and 1.7~h.

\section{Outlook}

Internal-conversion spectroscopy of $^{229}$Pa produced by electron capture of trapped $^{229}$U$^+$ in a Penning trap provides a direct route to searching for its candidate ground-state parity doublet, with sub-eV precision within about five days of measurement time for $\Delta E \gtrsim 20$~eV, and in less than two days for $\Delta E \gtrsim 60$~eV. In addition, the pattern of IC lines, fixed by known differences of ionization energies, provides a consistency check and a signature of the nuclear origin of the signal that is independent of $\Delta E$. Between 18.6~eV and the ionization energy of Pa$^{+}$, 11.9~eV, only the weak Pa$^{+}$ line is open. Because Pa$^+$ daughters arise almost exclusively from outer-shell capture, they carry neither prompt electrons nor K X rays, so the trigger described above is essentially blind to them. Such events can still be tagged by the $\gamma$ rays emitted by the nucleus as it cascades into the doublet state, which do not depend on the electronic relaxation. Requiring a detected $\gamma$ ray and no prompt electrons would select these events while rejecting most of the background. The achievable rate is set by the fraction of feeding transitions that proceed by $\gamma$ emission rather than internal conversion, since conversion would raise the charge state above Pa$^+$. Splittings below the ionization energy of Pa$^+$ are inaccessible to IC in the trapped ions in their electronic ground state, where the upper doublet member can decay only radiatively. For such splittings, techniques developed for the $^{229}$Th isomer can be adapted. For example, extracting the ions and neutralizing them in flight, opens IC down to the ionization energy of neutral Pa, $5.89$~eV, and allows the IC electrons to be analysed with a retarding-field spectrometer, as done in Ref.~\cite{Seiferle2019}.

A measurement of $\Delta E$ would remove the dominant uncertainty in the size of the $^{229}$Pa Schiff moment, which, for splittings of tens to hundreds of eV, would make this isotope the most sensitive nucleus to CP-violating phenomena~\cite{ArrowsmithKron2024,udrescu2026probing}. The proposed experiment would thereby establish whether $^{229}$Pa is a viable system for a new generation of searches for time-reversal violation in atoms, ions, and molecules. Moreover, the same platform can be readily extended to searches for low-lying nuclear states in other isotopes, which could play an important role in the field of fundamental symmetry violations, nuclear structure and nuclear clocks~\cite{udrescu2026probing}.

\section{Acknowledgments}
\label{sec:Ack}

S.-M.U. thanks David DeMille and Xing Fan for stimulating discussions. S.-M.U. was supported by a William H. Miller Fellowship. R.F.G.R. acknowledges support from the U.S. Department of Energy, Office of Science, Office of Nuclear Physics, under Grant No. DE-SC0021176, and from the Center for Ultracold Atoms (an NSF Physics Frontier Center).

Generative AI tools, including ChatGPT and Claude, were used during the preparation of this manuscript to assist with literature searches, manuscript organization, language editing, and figure editing.

\bibliographystyle{apsrev4-2}
\bibliography{references}

\newpage
\onecolumngrid
\newpage
\newcommand{\PRLsep}{\noindent\makebox[\linewidth]{\resizebox{0.75\linewidth}{1.75pt}{$\bullet$}}\bigskip}

\appendix
\begin{center}
    \vspace{0.5cm}
    \PRLsep
    \vspace{0.5cm}
    \textbf{\Large Supplemental Material}
\end{center}

\section{Transit-time tail of prompt low-energy electrons}
\label{sec:transit}

Each electron-capture (EC) decay produces, in addition to the prompt
trigger radiation, a continuum of low-energy electrons from the late
stages of the atomic vacancy cascade (outer-shell Auger and shake-off
electrons). Their energies overlap the expected range of the
internal-conversion (IC) electrons. Although these electrons are emitted
within $\lesssim 1$~ns of the decay, their arrival at the detector is
delayed by the transit through the retarding-field analyser: electrons
that clear the barrier with little excess energy cross it slowly. Here
we estimate how many such electrons arrive late enough to fall inside the
delayed-coincidence window. We show that this contribution is small
compared with the expected signal for delays $\gtrsim 1~\mu$s.

\paragraph{Longitudinal energy at the barrier.}
Consider an electron emitted with kinetic energy $E$ and pitch angle
$\theta$ relative to the magnetic field $B_0$ at the ion cloud. Under adiabatic transport, the orbital magnetic moment $\mu = E_\perp/B$ of the electron is conserved, so that its transverse kinetic energy decreases in proportion to the magnetic field. At the analysing
plane, the magnetic field is $B_A$ and the retarding potential energy
relative to the emission point is $eU$. There, the longitudinal kinetic
energy is
\begin{equation}
  E_\parallel = E - eU - E\sin^2\theta\,\frac{B_A}{B_0}.
  \label{eq:Epar}
\end{equation}
Electrons with $E_\parallel < 0$ are reflected; those with
$E_\parallel > 0$ are transmitted.

\paragraph{Transit time.}
For simplicity, we model the top of the barrier as approximately flat over an effective length $L$. An electron with longitudinal velocity
$v_\parallel = \sqrt{2E_\parallel/m_e}$ crosses this region in a time
$t_A \simeq L/v_\parallel$. Its transit therefore exceeds a time $t$
only if
\begin{equation}
  E_\parallel < E_t \equiv \frac{m_e L^2}{2t^2}
  \simeq 0.28~\text{meV}
  \left(\frac{L}{1~\text{cm}}\right)^{2}
  \left(\frac{1~\mu\text{s}}{t}\right)^{2}.
  \label{eq:Et}
\end{equation}
A delay of $1~\mu$s thus requires the electron to clear the barrier by less than $\sim 0.3$~meV.

\paragraph{Fraction of late prompt electrons.}
Let $\rho_0 \approx \left. dN_p/dE \right|_{E \approx eU}$ be the
spectral density of prompt low-energy electrons per decay near the
barrier energy. Calculated Auger spectra for the analogous EC decay of
$^{125}$I~\cite{Pomplun2000} give $\rho_0 \approx 0.05$~eV$^{-1}$ per decay
between 20 and 100~eV, corresponding to $4$ of the $5$
electrons per decay below 100~eV quoted in the main text. In Eq.~\eqref{eq:Epar}, $E_\parallel$ differs from $E$
only by the constant shift $eU$ and a perpendicular term suppressed by
$B_A/B_0$. The density of electrons near $E_\parallel = 0$ is therefore
also of order $\rho_0$. The number of prompt electrons detected per decay
with a transit delay exceeding $t$ is then
\begin{equation}
  P_p(>t) \approx \epsilon_e\,\rho_0\,E_t
  = \epsilon_e\,\rho_0\,\frac{m_e L^2}{2t^2},
  \label{eq:Pp}
\end{equation}
where $\epsilon_e$ is the electron detection efficiency. The cumulative
tail falls as $t^{-2}$. Thus, a tenfold increase in delay reduces the number of late prompt electrons a hundredfold. Using $\rho_0 \approx 0.05$ eV$^{-1}$, $\epsilon_e = 0.25$ and $L = 1$~cm, we obtain the values listed in Table~\ref{tab:tail}.

\begin{table}[ht]
\centering
\caption{
Estimated transit-time tail of prompt low-energy electrons for a flat barrier region of effective length $L=1$ cm, an assumed spectral density $\rho_0=0.05$ eV$^{-1}$ per decay, and electron detection efficiency $\epsilon_e=0.25$. Here $E_t=m_eL^2/(2t^2)$ and $P_{\rm p}(>t)=\epsilon_e\rho_0E_t$ is the mean number of detected prompt electrons per decay arriving after $t$ and $S_{IC}\simeq5.5\times10^{-4}$ at $\Delta E=60$ eV.
}
\label{tab:tail}
\begin{tabular}{cccc}
\hline\hline
Delay $t$ & $E_t$ (meV) &
$P_{\rm p}(>t)$ per decay & $P_{\rm p}/S_{IC}$ \\
\hline
$100$ ns
& $28$
& $3.6\times10^{-4}$
& $6.5\times10^{-1}$ \\

$500$ ns
& $1.1$
& $1.4\times10^{-5}$
& $2.6\times10^{-2}$ \\

$1\,\mu{\rm s}$
& $0.28$
& $3.6\times10^{-6}$
& $6.5\times10^{-3}$ \\

$5\,\mu{\rm s}$
& $0.011$
& $1.4\times10^{-7}$
& $2.6\times10^{-4}$ \\
\hline\hline
\end{tabular}
\end{table}

\paragraph{Comparison with the signal.}
The expected number of detected IC electrons per trigger is $S_{IC} \approx p_\text{pop} f_\text{IC}^\text{trig} \epsilon_e$, where $p_\text{pop} \approx 0.1$ is the probability of populating the candidate state per EC decay and $f_\text{IC}^\text{trig}$ is the fraction of triggered daughters for which IC is energetically open. For $\Delta E = 60$~eV, the calculated Pa fraction $f_\text{IC} \approx 0.02654$ of the main text, together with trigger probabilities of $\approx 0.821$ for IC-open daughters and $\approx 0.992$ for all decays, gives $f_\text{IC}^\text{trig} \approx 0.0220$, so $S_{IC} \approx 5.5\times10^{-4}$ per trigger. The late prompt electrons then amount to $\approx 0.65\%$ of the signal for delays above 1~$\mu$s. We therefore open the coincidence window at $t_0 \approx 1$~$\mu$s, and include any residual tail as a separate component when fitting the delayed-electron time spectra.

\paragraph{IC electrons versus  prompt electrons.}
At a given $E_\parallel$, IC electrons experience the same transit-time response as prompt electrons. Their arrival-time distribution is that response convolved with the exponential decay of the excited parity-doublet state. The difference is that IC electrons form discrete lines. Unless the barrier is set within $\sim E_t$ of a line, which happens only in the step region, they clear the barrier with an excess energy of order eV and cross it in at most a few tens of nanoseconds. The prompt continuum, by contrast, always contains electrons just above threshold, irrespective of the retarding setting. These electrons produce the tail estimated above.

\paragraph{Validity and direct measurement of the response.} Equation~\eqref{eq:Pp} is conservative with respect to the barrier shape. Consider a smooth maximum of the electron potential energy along the axis, $V(z) \simeq eU - \tfrac{1}{2}k z^2$. The time spent near the top then grows only logarithmically as $E_\parallel \to 0$: $t_A \sim \omega^{-1}\ln(\Delta/E_\parallel)$, with $\omega = \sqrt{k/m_e}$ and $\Delta$ the energy scale of the barrier top. The tail then falls exponentially rather than as a power law.

The estimate scales linearly with $\rho_0$ and quadratically with $L$. A denser prompt spectrum near the analysed energies, or a longer flat region in the analyser, would increase the tail correspondingly. The analysis will therefore not rely on Eq.~\eqref{eq:Pp}. Instead, we will measure the response directly in two ways: (i) at retarding settings above the highest IC line, where only prompt and uncorrelated electrons are transmitted; and (ii) with pulsed calibration electrons of known energy.

\section{Adaptive determination of $\Delta E$}
\label{app:fit}

We describe here the analysis used to choose the barrier settings and to extract $\Delta E$, and the simulations behind the measurement times quoted in the main text. The analysis does not assume known signal and background rates.

\textit{Expected counts.} Each delayed count is assigned to one of four multiplicity classes according to the number of prompt electrons detected in its trigger: $c=0$ (trigger from the K X ray alone), 1, 2, or $\ge3$. For a cycle of live time $t$ at a barrier $U$, held fixed during the cycle, the expected number of counts in class $c$ is
\begin{equation}
\mu_c(U)=t\Big[A\sum_{q\ge2}R_q\,f_{qc}\,T(\Delta E-I_q,U)+B\,g_c\Big].
\label{eq:mu}
\end{equation}
Here $R_q=(p_{\rm pop}N_{\rm EC}/t_c)\,P(q)\,\eta_q\,\epsilon_e$ is the nominal height of the Pa$^{q+}$ step, with $p_{\rm pop}=0.1$, $N_{\rm EC}=1.9\times10^3$ decays per measurement cycle of length $t_c=20$~min (20 trap cycles of 1 min), $\epsilon_e=0.25$ is the probability of detecting an IC electron, $\eta_q$ is the prompt trigger probability, defined in the main text, and $P(q)$ is the probability distribution of charge states of the Pa daughter from Fig.~\ref{figure_charge_states}; Pa$^+$ is untagged. $A$ (nominally 1) and $B$ (nominally 1.24~counts/day) are the unknown signal scale and background rate. For an electron of kinetic energy $K$ emitted isotropically into the analysed hemisphere, Eq.~(\ref{eq:Epar}) gives the transmission
\begin{equation}
T(K,U)=\int_0^1\!d\mu\;\Phi\!\left(\frac{K\,[1-\beta(1-\mu^2)]-eU}{\sigma_r}\right),
\label{eq:T}
\end{equation}
with $\mu=\cos\theta$, $\beta=B_A/B_0=0.01$, $\Phi$ the standard normal cumulative distribution, and $\sigma_r=0.1$~eV a Gaussian energy spread; $T=0$ for $K\le0$. The fraction $f_{qc}$ of triggered Pa$^{q+}$ events in class $c$ assumes that each of the $q-1$ prompt electrons is detected independently with probability $p_e=0.5$ and the K X ray with $p_X=0.15$: $f_{q0}\propto p_X(1-p_e)^{q-1}$ and $f_{qk}\propto\binom{q-1}{k}p_e^k(1-p_e)^{q-1-k}$ for $k\ge1$ (summed over $k\ge3$), normalized by $\eta_q$. Background triggers have the multiplicity distribution $g_c$ of all EC decays, dominated by high charge states. Table~\ref{tab:inputs} lists the inputs.

\begin{table}[t]
\caption{Inputs to Eq.~(\ref{eq:mu}). Ionization energies $I_q$ from Ref.~\cite{NISTASD}; step heights $R_q$ for $A = 1$ ($P(q)$ for $q \ge 5$ taken from the MCRE calculation, Fig.~\ref{figure_charge_states});
fractions $f_{qc}$ of triggered events with $c = 0$ (K X ray only), 1, 2, and ${\ge}3$ detected prompt electrons. Last row: multiplicity
distribution $g_c$ of the background.}
\label{tab:inputs}
\begin{ruledtabular}
\begin{tabular}{lcccccc}
$q$ & $I_q$ (eV) & $R_q$ (counts/h) & $c=0$ & $c=1$ & $c=2$ & $c\ge3$ \\
\hline
2 & 18.6 & 0.244 & 0.130 & 0.870 & 0 & 0 \\
3 & 30.9 & 0.820 & 0.048 & 0.635 & 0.317 & 0 \\
4 & 44.3 & 2.05 & 0.021 & 0.420 & 0.420 & 0.140 \\
5 & 72.0 & 2.17 & 0.010 & 0.264 & 0.396 & 0.330 \\
6 & 85.1 & 4.18 & 0.005 & 0.161 & 0.321 & 0.514 \\
7 & 98.9 & 7.48 & 0.002 & 0.095 & 0.238 & 0.665 \\
8 & 111.0 & 9.24 & 0.001 & 0.055 & 0.165 & 0.779 \\
\hline
\multicolumn{3}{l}{Background, $g_c$} & 0.001 & 0.035 & 0.071 & 0.892 \\
\end{tabular}
\end{ruledtabular}
\end{table}

\textit{Priors and posterior.} For this simulation we focus on the energy range of $20-110$ eV, but extending it to higher energies is straightforward. We assume that $\Delta E$ has a uniform prior over 18.7--112~eV, slightly wider than the range of interest to avoid edge effects; $\ln A$ is Gaussian with mean 0 and width $\ln3$, and $\ln B$ is Gaussian about the expected background with width 0.3. The ionization energies are fixed at the tabulated values of Table~\ref{tab:inputs}. Let $N_{ic}$ be the number of delayed counts recorded in class $c$ during cycle $i$. For given parameters it is Poisson distributed with mean $\mu_{ic}=\mu_c(U_i)$ from Eq.~(\ref{eq:mu}), so the likelihood of the data is
\begin{equation}
L=\prod_{i}\prod_{c}\frac{\mu_{ic}^{\,N_{ic}}\,e^{-\mu_{ic}}}{N_{ic}!}.
\label{eq:L}
\end{equation}
The posterior $p(\Delta E,A,B\,|\,\text{data})$ is computed numerically on a grid: $\Delta E$ in 0.02-eV steps over the full range and, for each $\Delta E$, $40\times40$ values of $\ln A$ and $\ln B$, whose range is narrowed to the region of non-negligible probability as data accumulate. Summing over $A$ and $B$ gives the posterior of $\Delta E$ alone, from which the mean, the 68\% interval, and $P_{\rm main}$ are obtained (see below).

\textit{Choice of the next setting.} The first cycle is taken at $U=0$. Each subsequent setting maximizes the expected information gain on $\Delta E$, i.e., the mutual information between $\Delta E$ and the counts of the next cycle~\cite{Lindley1956},
\begin{equation}
\mathcal I(U)=\sum_c\Big\{H\big[p(N_c|U)\big]-\sum_kP_k\,H\big[p(N_c|U,k)\big]\Big\},
\label{eq:gain}
\end{equation}
where $H[p]=-\sum_Np(N)\ln p(N)$, $k$ labels bins of $\Delta E$, defined below, with current posterior probabilities $P_k$, $p(N_c|U,k)$ is the predicted count distribution in bin $k$, and $p(N_c|U)=\sum_kP_k\,p(N_c|U,k)$. As long as the 68\% half-width exceeds the 1~eV target, the bins are 0.5~eV wide, aligned so that the posterior mode lies at a bin centre, and the chosen setting is the one that best narrows $\Delta E$. Once the target is met, the bins are widened to 6~eV, so that the main peak (mode $\pm3$~eV) forms a single bin; the gain then rewards only settings that discriminate the main peak from competing charge-state assignments, the remaining requirement of the stopping rule. Candidate settings form a 0.25-V grid from 0 to 100~V, refined to 0.05~V near the predicted step edges. The choice of setting affects only the efficiency; the posterior always uses the full likelihood.

\textit{Stopping rule.} Let $M$ be the value of $\Delta E$ at which the posterior is largest. The measurement stops when $P_{\rm main}\equiv P(|\Delta E-M|\le3~\text{eV})\ge0.99$ and the central 68\% credible interval has a half-width $\le1$~eV, in two consecutive cycles. Alternative charge-state assignments differ by at least 12~eV, so the first condition requires the posterior probability of a misassignment to be  $\le 1\%$, and the confirmation cycle guards against a single fluctuation. $\Delta E$ is reported as the posterior mean with its 68\% interval. 

\textit{Simulations.} Simulated experiments use $A=1$ and $B=1.24$~counts/day, independent of $U$, and the true $\Delta E$ drawn within $\pm0.1$~eV ($\pm0.5$~eV at 60~eV) of the nominal value; the analysis is given none of these. For each experiment, we record the live time at which the stopping rule is met and the error of the reported value, i.e., its difference from the true $\Delta E$. Each experiment is then continued until a stricter rule, a 68\% half-width $\le0.5$~eV and $P_{\rm main}\ge0.999$, is met. Table~\ref{tab:sim} summarizes the results. The 68\% intervals contain the true value in 64\% of the 90 experiments, close to the nominal 68\%. None of the experiments ended on a wrong charge-state assignment.

\begin{table}[t]
\caption{Simulated performance: number of experiments; mean live time until the stopping rule (${\pm}1$~eV, $P_\text{main} \ge 0.99$)
is met; root-mean-square difference between the reported and true $\Delta E$ at that time; mean live time until the stricter rule
(${\pm}0.5$~eV, $P_\text{main} \ge 0.999$) is met; experiments ending on a wrong charge-state assignment.}
\label{tab:sim}
\begin{ruledtabular}
\begin{tabular}{cccccc}
$\Delta E$ (eV) & $N_\text{exp}$ & Time to ${\pm}1$~eV (h) & RMS error (eV) & Time to ${\pm}0.5$~eV (h) & Wrong \\
\hline
20 & 20 & 117.4 & 0.31 & 165.5 & 0 \\
60 & 40 & 31.4 & 0.35 & 41.1 & 0 \\
110 & 30 & 8.7 & 0.39 & 12.0 & 0 \\
\end{tabular}
\end{ruledtabular}
\end{table}

\section{Charge-state distribution of $^{229}\mathrm{Pa}$}

\begin{figure*}
    \centering
    \includegraphics[width=\textwidth]{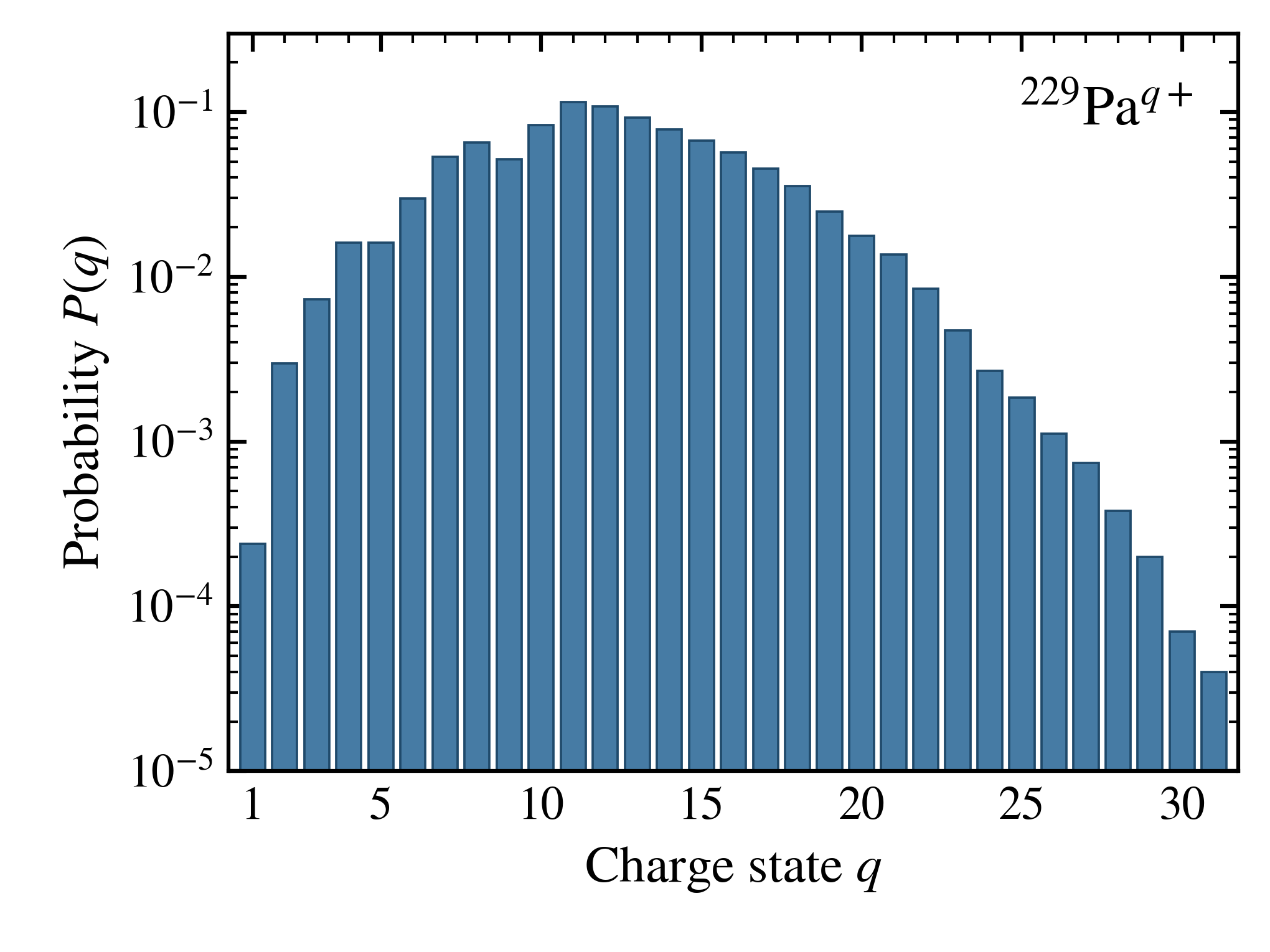}
    \caption{\textbf{Charge-state distribution of $^{229}\mathrm{Pa}$.} Bars show the probability $P(q)$ for each ionic charge state $^{229}\mathrm{Pa}^{q+}$. The vertical axis is logarithmic. The distribution peaks at $q=11$, with a probability of approximately $11.5\%$.}
    \label{figure_charge_states}
\end{figure*}

The charge-state distribution of $^{229}$Pa following electron capture (EC) in $^{229}$U in Fig. \ref{figure_charge_states} was calculated using our previously developed Monte Carlo rate-equation (MCRE) framework \cite{Ho2014}. MCRE tracks the evolution of electronic configurations by stochastically sampling competing atomic transitions, enabling efficient simulation of complex multiple-ionization cascades.  The calculation was initialized with a single $1s_{1/2}$ (K-shell) vacancy in neutral Pa ($Z=91$). Subsequent relaxation through radiative, Auger, Coster-Kronig (CK), and super-Coster-Kronig (sCK) transitions was simulated using atomic rates calculated with the Flexible Atomic Code (FAC) \cite{Gu2008}.  At each Monte Carlo step, an energetically allowed transition was selected with probability proportional to its rate. Radiative transitions redistribute vacancies without changing the charge state, whereas nonradiative transitions increase the ionic charge. Each trajectory was followed until no further energetically allowed transitions remained.  The final charge-state distribution was obtained from $10^5$ independent Monte Carlo trajectories.

\end{document}